\documentclass[preprintnumbers,superscriptaddress,showpacs,pra]{revtex4}%
\usepackage[colorlinks=true,linkcolor=blue]{hyperref}
\usepackage{amssymb}
\usepackage{amsfonts}
\usepackage{amsmath}
\usepackage{graphicx}%
\providecommand{\U}[1]{\protect\rule{.1in}{.1in}}

\let\stdsection\section
\renewcommand\section{\nopagebreak\stdsection}
\begin{document}
\preprint{REV\TeX4-1}
\title{A self-adjoint decomposition of radial momentum implies that Dirac's
introduction of the operator is insightful}
\author{Q. H. Liu}
\email{quanhuiliu@gmail.com}
\affiliation{School for Theoretical Physics, and Department of Applied Physics, Hunan
University, Changsha, 410082, China}
\author{S. F. Xiao}
\affiliation{School for Theoretical Physics, and Department of Applied Physics, Hunan
University, Changsha, 410082, China}
\affiliation{Department of Physics, Zhanjiang Normal University, Zhanjiang 524048, China}
\date{\today}

\begin{abstract}
With acceptance of the Dirac's observation that the \textit{canonical
quantization entails using Cartesian coordinates, }we examine the\textit{\ }%
operator $\mathbf{e}_{r}P_{r}$ rather than $P_{r}$ itself and demonstrate that
there is a decomposition of $\mathbf{e}_{r}P_{r}$ into two self-adjoint but
non-commutative parts, in which one is the total momentum and another is the
transverse one. This study renders the operator $P_{r}$ indirectly measurable
and physically meaningful.

\end{abstract}

\maketitle

\section{Introduction}

It is a puzzle that Dirac simply overlooked the claim that the radial momentum
$P_{r}=-i\hbar(\partial_{r}+1/r)$ is not self-adjoint and "is not the
$r$-component of the particle momentum" which was explicitly demonstrated in
1960 \cite{DP} and since then \cite{RH,1968,LNF,DC,1989,2002}, and he insists
his own understanding that $P_{r}$ "is real and is the true momentum conjugate
to $r$", \cite{dirac} which is reflected in the last revision of his classic
book on quantum mechanics that published in 1967. On one hand, in an
apparently mathematical sense, Dirac was "wrong" because $P_{r}$ can never be
a self-adjoint operator. On the other, the mean values of the $P_{r}$ and its
second power $P_{r}^{2}$ over the state functions of, e.g., Hydrogen atom,
make sense, \cite{kuo,dong} and the underlying quantum principle states that
in a measurement of an observable an eigenvalue of the observable is obtained.
In other words, if the operator does not possess a complete set of the
eigenfunctions at all, to talk about the obtained value would be nonsense thus
its mean value is totally meaningless. Even though the converse of this
statement may not be necessarily true, it still must be instructive if one
tries to add some insight into the understanding of this operator $P_{r}$,
rather than simply to put it aside because of its non-self-adjointness. We
show that the radial momentum $P_{r}$ can be decomposed into a difference of
two noncommutative but self-adjoint operators each having well-defined eigenstates.

This paper is organized in the following. In section II, how to decompose the
radial momentum operator into two self-adjoint operator parts is given. In
section III, how to measure the radial momentum operator is presented. Section
IV is a brief conclusion and discussion.

\section{A decomposition of the radial momentum operator into two self-adjoint
operator parts}

The relation between the Cartesian coordinates ($x,y,z$) and the spherical
polar coordinates ($r,\theta,\varphi$) is,
\begin{equation}
x=r\sin\theta\cos\varphi,\text{ }y=r\sin\theta\sin\varphi,\text{ }%
z=r\cos\theta,
\end{equation}
where $r$ is the distance from the nucleus (origin) to the electron, and
$\theta$ is the polar angle from the positive $z$-axis with $0\leq\theta
\leq\pi$, and $\varphi$ is the azimuthal angle in the $xy$-plane from the
$x$-axis with $0\leq\varphi\leq2\pi$. The gradient operator in the Cartesian
coordinate system $\nabla_{cart}\equiv\mathbf{e}_{x}\partial_{x}%
+\mathbf{e}_{y}\partial_{y}+\mathbf{e}_{z}\partial_{z}$\ can also be expressed
in ($r,\theta,\varphi$),
\begin{equation}
\nabla_{sp}=\mathbf{e}_{r}\frac{\partial}{\partial r}+\mathbf{e}_{\theta}%
\frac{1}{r}\frac{\partial}{\partial\theta}+\mathbf{e}_{\varphi}\frac{1}%
{r\sin\theta}\frac{\partial}{\partial\varphi}. \label{grad}%
\end{equation}
It can be rewritten into%
\begin{equation}
\nabla_{sp}=\mathbf{e}_{r}(\frac{\partial}{\partial r}+\frac{1}{r}%
)+\nabla_{tran}, \label{decomp}%
\end{equation}
where $\nabla_{tran}$ is the transverse component of the gradient operator
$\nabla_{sp}$,
\begin{equation}
\nabla_{tran}=\mathbf{e}_{\theta}\frac{1}{r}\frac{\partial}{\partial\theta
}+\mathbf{e}_{\varphi}\frac{1}{r\sin\theta}\frac{\partial}{\partial\varphi
}-\mathbf{e}_{r}\frac{1}{r}. \label{tran}%
\end{equation}
By the transverse, we mean that it is perpendicular to the radial direction
$\mathbf{e}_{r}$, with considering the noncommutability between $\mathbf{e}%
_{r}$ and $\nabla_{tran}$,
\begin{equation}
\mathbf{e}_{r}\cdot\nabla_{tran}+\nabla_{tran}\cdot\mathbf{e}_{r}=0.
\label{otho}%
\end{equation}
Multiplied $\nabla_{tran}$ by a factor $-i\hbar$, we obtain the transverse
momentum $\mathbf{\Pi}$ defined by,%
\begin{equation}
\mathbf{\Pi=}-i\hbar\nabla_{tran}. \label{geom}%
\end{equation}
We have evidently $\mathbf{e}_{r}\cdot\mathbf{\Pi}+\mathbf{\Pi}\cdot
\mathbf{e}_{r}=0$. Noting that the momentum operator can be also written in
the following way,
\begin{align}
\mathbf{P}  &  \mathbf{\equiv}-i\hbar\nabla_{cart}=-i\hbar\nabla_{sp}\\
&  =\left\{  \mathbf{e}_{r},P_{r}\right\}  +\frac{1}{r}\left\{  \mathbf{e}%
_{\theta},P_{\theta}\right\}  +\frac{1}{r\sin\theta}\left\{  \mathbf{e}%
_{\varphi},P_{\varphi}\right\}  ,
\end{align}
where $\{A,B\}\equiv(AB+BA)/2$ and,
\begin{align}
\left\{  \mathbf{e}_{r},P_{r}\right\}   &  =\mathbf{e}_{r}P_{r},P_{r}%
=-i\hbar(\frac{\partial}{\partial r}+\frac{1}{r}),\\
\{\mathbf{e}_{\theta},P_{\theta}\}  &  =\mathbf{e}_{\theta}P_{\theta}%
+i\hbar\frac{\mathbf{e}_{r}}{2},P_{\theta}=-i\hbar(\frac{\partial}%
{\partial\theta}+\frac{\sin\theta}{\cos\theta}),\\
\{\mathbf{e}_{\varphi},P_{\varphi}\}  &  =\mathbf{e}_{\varphi}P_{\varphi
}+i\hbar\frac{1}{2}\left(  \mathbf{e}_{r}\sin\theta+\mathbf{e}_{\theta}%
\cos\theta\right)  ,P_{\varphi}=-i\hbar\frac{\partial}{\partial\varphi}.
\end{align}
We have then,%
\begin{equation}
\mathbf{\Pi}=\frac{1}{r}\left\{  \mathbf{e}_{\theta},P_{\theta}\right\}
+\frac{1}{r\sin\theta}\left\{  \mathbf{e}_{\varphi},P_{\varphi}\right\}  ,
\end{equation}
whose second decomposition in Cartesian coordinates gives,%
\begin{align}
\Pi_{x}  &  =-\frac{i\hbar}{r}(\cos\theta\cos\varphi\frac{\partial}%
{\partial\theta}-\frac{\sin\varphi}{\sin\theta}\frac{\partial}{\partial
\varphi}-\sin\theta\cos\varphi),\label{gmx}\\
\Pi_{y}  &  =-\frac{i\hbar}{r}(\cos\theta\sin\varphi\frac{\partial}%
{\partial\theta}+\frac{\cos\varphi}{\sin\theta}\frac{\partial}{\partial
\varphi}-\sin\theta\sin\varphi),\label{gmy}\\
\Pi_{z}  &  =\frac{i\hbar}{r}(\sin\theta\frac{\partial}{\partial\theta}%
+\cos\theta). \label{gmz}%
\end{align}
They are three components of the so-called geometric momentum for a particle
constrained on the two-dimensional spherical surface of radius $r$,
\cite{liu13-1,liu11,liu13-2} and each possesses complete set of the
eigenfunctions. In order to see it, let us consider following commutation
relations that form an $so(3,1)$ algebra,%
\begin{equation}
\lbrack r\Pi_{{i}}{,}r\Pi_{j}]=-i\hbar\varepsilon_{ijk}L_{k}\text{, }[{L}%
_{{i}}{,}r\Pi_{j}]=i\hbar\varepsilon_{ijk}r\Pi_{{k}}\text{, }[{L}_{{i}}%
{,L}_{j}]={i}\hbar\varepsilon_{ijk}L_{k}, \label{so31}%
\end{equation}
where $\boldsymbol{L}$ is the orbital angular momentum whose three Cartesian
components are,%
\begin{align}
{L_{x}}  &  =i\hbar(\sin\varphi\frac{\partial}{\partial\theta}+\cot\theta
\cos\varphi\frac{\partial}{\partial\varphi}),\\
{{L}_{y}}  &  =-i\hbar(\cos\varphi\frac{\partial}{\partial\theta}-\cot
\theta\sin\varphi\frac{\partial}{\partial\varphi}),\\
{{L}_{z}}  &  =-i\hbar\frac{\partial}{\partial\varphi}.
\end{align}
One can easily verify that $r\mathbf{\Pi}$ is the real part of the operator
$\boldsymbol{L\times}\mathbf{e}_{r}$, \cite{liu13-1}
\[
\boldsymbol{L\times}\mathbf{e}_{r}=r\mathbf{\Pi+}i\hbar\mathbf{e}_{r}.
\]
Three commutable pairs (${L}_{{i}}{,}r\Pi_{{i}}$) are equivalent with each
other upon a rotation of the coordinate system, \cite{liu11,liu13-2}
\begin{equation}
f_{x}=\exp(-i\pi L_{y}/2)f_{z}\exp(i\pi L_{y}/2),\text{ }f_{y}=\exp(i\pi
L_{x}/2)f_{z}\exp(-i\pi L_{x}/2),\text{ }(f_{i}\rightarrow L_{i}\text{ or
}r\Pi_{{i}}). \label{rotation}%
\end{equation}
Here we follow the convention that a rotation operation affects a physical
system itself. \cite{Sakurai} Equation (\ref{rotation}) above implies that it
is sufficient to study one representation determined by one pair of the three
$[{L}_{{i}}{,}\Pi_{{i}}]=0$. For simplicity, we have the complete set of three
commuting observables ($r,\Pi_{{z}},L_{z}$) or ($r,r\Pi_{{z}},L_{z}$), and it
is evident for they are independent from each other. The normalized
eigenfunctions are, respectively,
\begin{align}
r\delta(r-r^{\prime})  &  =r^{\prime}\delta(r-r^{\prime}),\label{rr}\\
r\Pi_{{z}}\frac{1}{\sqrt{2\pi}}\frac{1}{\sin\theta}{\exp\left(  -i\gamma{_{z}%
}\ln\tan\frac{\theta}{2}\right)  }  &  =\gamma{_{z}}\hbar\frac{1}{\sqrt{2\pi}%
}\frac{1}{\sin\theta}{\exp\left(  -i{{\gamma}_{z}}\ln\tan\frac{\theta}%
{2}\right)  ,}\label{rgeom}\\
L_{z}\frac{1}{\sqrt{2\pi}}e^{im\varphi}  &  =m\hbar\frac{1}{\sqrt{2\pi}%
}e^{im\varphi}, \label{lz}%
\end{align}
where $\delta(r-r^{\prime})$ is delta-function normalized eigenfunction of
radius $r$ with respect to the measure $dr$ rather than $r^{2}dr$ with which
$\delta(r-r^{\prime})$ should be replaced by $\delta(r-r^{\prime})/r$, and
$\gamma{_{z}\hbar\in(-\infty,\infty)}$ is the eigenvalue of operator
$r\Pi_{{z}}$, and other symbols have their usual meanings. The Eq.
(\ref{rgeom}) can be rewritten into following form,%
\begin{equation}
\Pi_{{z}}\delta(r-r^{\prime})\frac{1}{\sqrt{2\pi}}\frac{1}{\sin\theta}%
{\exp\left(  -i\gamma{_{z}}\ln\tan\frac{\theta}{2}\right)  =}\frac{\gamma
{_{z}}}{r^{\prime}}\delta(r-r^{\prime})\frac{1}{\sqrt{2\pi}}\frac{1}%
{\sin\theta}{\exp\left(  -i\gamma{_{z}}\ln\tan\frac{\theta}{2}\right)  .}
\label{rgeom1}%
\end{equation}

Thus the radial momentum $\mathbf{e}_{r}P_{r}$ can be defined as the
difference of the total one $-i\hbar\nabla_{cart}$ and the transverse part
$\mathbf{\Pi}$,%
\begin{equation}
\mathbf{e}_{r}P_{r}\equiv-i\hbar\nabla_{cart}-\mathbf{\Pi.} \label{def}%
\end{equation}
Two operators $-i\hbar\nabla_{cart}$ and $\mathbf{\Pi}$ are self-adjoint
respectively, and it is evidently $\left[  -i\hbar\nabla_{cart},\mathbf{\Pi
}\right]  \neq0$. The former has the eigenfunctions $\{\psi_{p_{x}}%
(x)\psi_{p_{y}}(y)\psi_{p_{z}}(z)\}$ with $\psi_{p_{i}}(x)=(\sqrt{2\pi\hbar
})^{-1}\exp(ix_{i}p_{i}/\hbar)$ ($i=1,2,3$) whereas the latter has the eigenfunctions

\{$\delta(r-r^{\prime})(\sqrt{2\pi}\sin\theta)^{-1}{\exp\left(  -i{{\gamma
}_{z}}\ln\tan(\theta/2)\right)  }(\sqrt{2\pi})^{-1}e^{im\varphi}$\}.

The square of the quantity $\left(  \mathbf{e}_{r}P_{r}\right)  ^{2}$ is
simply from (\ref{def}),%
\begin{align}
P_{r}^{2}  &  =\sum_{j=1}^{3}\left(  i\hbar\partial_{j}+\Pi_{j}\right)
\left(  i\hbar\partial_{j}+\Pi_{j}\right) \nonumber\\
&  =-\hbar^{2}\nabla_{cart}^{2}-\frac{L^{2}}{r^{2}} \label{pr2}%
\end{align}
This is also the difference of the square of total momentum $-\hbar^{2}%
\nabla_{cart}^{2}$ and the transverse part $L^{2}/r^{2}$.

\section{On the measurement of the radial momentum operator $\mathbf{e}%
_{r}P_{r}$}

Spherical polar coordinates are practically useful for a system where the
separation of variables in these coordinates is possible. We deal with only
stationary state of form $R_{nl}(r)Y_{lm}(\theta,\varphi)$ where the measure
is $r^{2}dr$, and if the measure is $dr$ the state function is $rR_{nl}%
(r)Y_{lm}(\theta,\varphi)$ instead. For a measurement of the quantity
$\mathbf{e}_{r}P_{r}$, we need to measure $-i\hbar\nabla_{cart}$ and
$\mathbf{\Pi}$, separately, on the same prepared state, so they are two with
interference-free measurements.

In order to measure the quantity $\mathbf{e}_{r}P_{r}$, we need to measure
three Cartesian components respectively, denoted by ($P_{rx},P_{ry},P_{rz}$).
And, the measurement of the operator $\mathbf{\Pi}$ means that for a given
radius $r$ we measure a pair of commutating quantities $[\Pi_{i},L_{i}]$
($i=1,2,3$). Because the probability of the particle located at the given
radius interval $dr$ is proportional to $\left\vert rR_{nl}(r)\right\vert
^{2}dr$, obtained values of the quantities $[\Pi_{i},L_{i}]$ must be
integrated over $r\in\lbrack0,\infty)$ in order to get pure transverse
component of the total momentum.

For a stationary state $\psi_{nlm}(r,\theta,\varphi)=R_{nl}(r)Y_{lm}%
(\theta,\varphi)$, \cite{Schiff} we have its expansion in the ($r,\Pi_{{z}%
},L_{z}$) representation,%
\begin{equation}
R_{nl}(r)Y_{lm}(\theta,\varphi)=R_{nl}(r)\sum_{m^{\prime}}\delta_{m^{\prime}%
m}{\frac{1}{\sqrt{2\pi}}e^{im^{\prime}\varphi}}\int Q_{lm}({{\gamma}_{z}%
})\frac{1}{\sqrt{2\pi}}\frac{1}{\sin\theta}{\exp\left(  -i{{\gamma}_{z}}%
\ln\tan\frac{\theta}{2}\right)  d{\gamma}_{z},}%
\end{equation}
where,
\begin{equation}
Q_{lm}({{\gamma}_{z}})=\oint Y_{lm}(\theta,\varphi)\frac{1}{\sqrt{2\pi}}%
\frac{1}{\sin\theta}{\exp\left(  i{{\gamma}_{z}}\ln\tan\frac{\theta}%
{2}\right)  \sin\theta d\theta d\varphi.} \label{qlm}%
\end{equation}
The explicit form of the coefficients $Q_{lm}({{\gamma}_{z}})$ is available
from literature. \cite{liu13-2} The value of $Q_{00}({{\gamma}_{z}})$ is
easily carried out, which is,
\[
Q_{00}({{\gamma}_{z}})=\frac{1}{2}\sqrt{\pi}\text{sech}\left(  \frac{\pi}%
{2}{{\gamma}_{z}}\right)  .
\]

In the following, we present the distribution of the $z$-axis component of
$\mathbf{e}_{r}P_{r}$ for the ground state of the Hydrogen atom. The ground
state of the Hydrogen atom is,%
\begin{equation}
\psi_{100}=2\left(  \frac{1}{a_{0}}\right)  ^{3/2}\exp\left(  -\frac{r}{a_{0}%
}\right)  \frac{1}{\sqrt{4\pi}}, \label{ground}%
\end{equation}
where $a_{0}$ is the Bohr radius, and it is isotropic. The momentum
distribution amplitude is,
\begin{align}
c\left(  p_{x},p_{y},p_{z}\right)   &  =\int\psi_{100}\exp(-\frac{i}{\hbar
}pr\cos\theta)r^{2}dr{\sin\theta d\theta d\varphi}\\
&  =\frac{2^{3/2}b^{5/2}}{\pi}\frac{1}{(p^{2}+b^{2})^{2}},
\end{align}
where $b\equiv\hbar/a_{0}$, and probability in interval $dp_{x}dp_{y}dp_{z}$
is given by,%
\begin{equation}
\left\vert c\left(  p_{x},p_{y},p_{z}\right)  \right\vert ^{2}=\frac
{2^{3}b^{5}}{\pi^{2}}\frac{1}{(p^{2}+b^{2})^{4}}.
\end{equation}
We are only interested in the distribution along $z$-axis, which is given by,%
\begin{align}
\left\vert c\left(  p_{z}\right)  \right\vert ^{2}  &  =\int\left\vert
c\left(  p_{x},p_{y},p_{z}\right)  \right\vert ^{2}dp_{x}dp_{y}\nonumber\\
&  =\frac{8b^{5}}{3\pi}\frac{1}{(b^{2}+p_{z}^{2})^{3}}.
\end{align}

Next, we consider the $z$-axis component of the transverse momentum $\Pi_{z}$.
The distribution in interval $\hbar d{{\gamma}_{z}}$ is nothing but
$\left\vert Q_{00}({{\gamma}_{z}})\right\vert ^{2}/a_{0}$, where $1/a_{0}$
comes from mean value of $1/r$ factor of the operator $\Pi_{z}$ (\ref{gmz}) in
the state (\ref{ground}). The curves of momentum $\left\vert c\left(
p_{z}\right)  \right\vert ^{2}$ and $\left\vert Q_{00}({{\gamma}_{z}%
})\right\vert ^{2}/a_{0}$ are plotted in Fig. 1 with the same parameter
${{\gamma}_{z}}$ characterizing the magnitude of the $z$-axis component the momentum.


\begin{figure}
[htb]%
\includegraphics[width=0.9\textwidth]{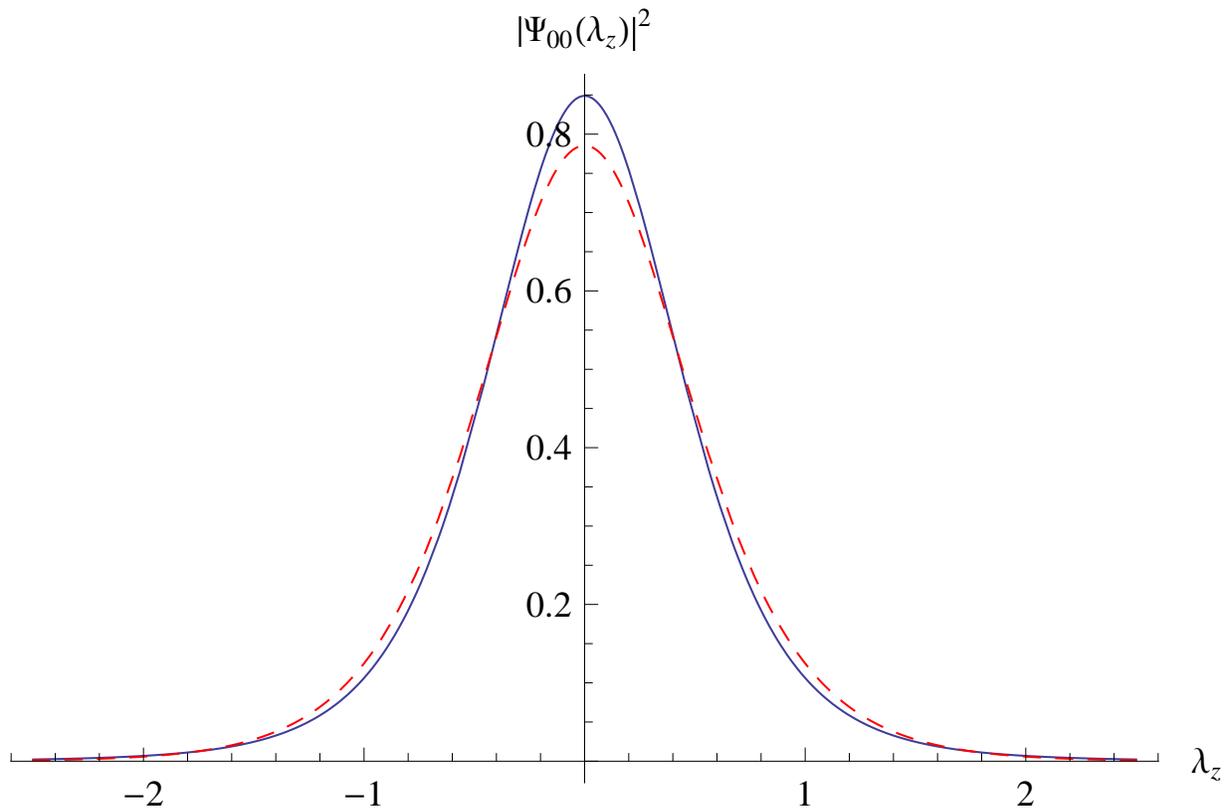}\caption{Distribution densities of $p_{z}$ (solid line) and  $\Pi_{z}$ (dashed line)
versus momentum magnitude $\lambda_z \hbar/a_0$ for the ground state of Hydrogen atom
$\psi_{100}$. In figures, natural units $\hbar=a_0=1$ are used.}\label{figure 1}%

\end{figure}

\begin{figure}
[htb]%
\includegraphics[width=0.9\textwidth]{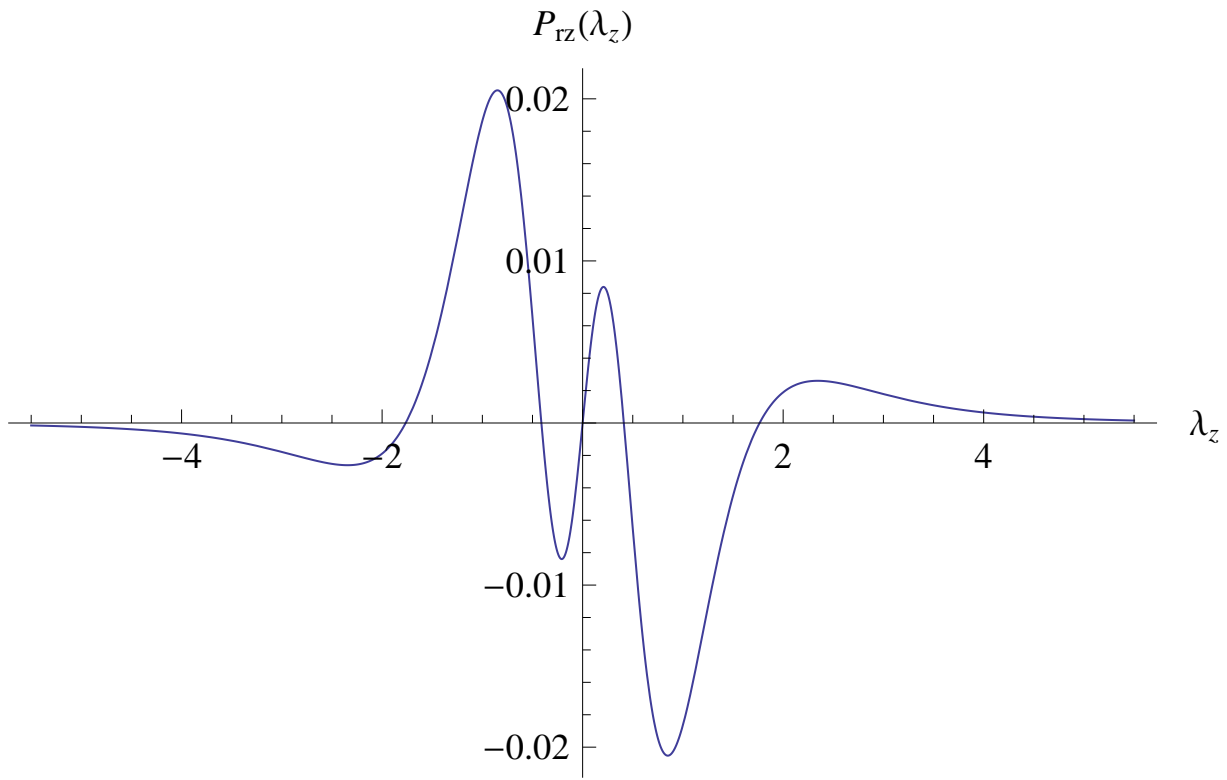}\caption{Distribution of the $z$-component of $e_{r}P_{r}$ against the
dimensionless momentum magnitude $\lambda_z$ for the ground state of Hydrogen atom.
The mean value is zero because of the antisymmetry about $\lambda_z=0$.}\label{figure 2}%

\end{figure}


Now, we are in position to give the distribution of the $z$-axis component of
the quantity $\mathbf{e}_{r}P_{r}$ in terms of the parameter ${{\gamma}_{z}}$.
It is clearly given by,%
\begin{equation}
{{\gamma}_{z}}\left[  \frac{8}{3\pi}\frac{1}{(1+{{\gamma}_{z}^{2}})^{3}}%
-\frac{\pi}{4}\text{sech}^{2}\left(  \frac{\pi}{2}{{\gamma}_{z}}\right)
\right]  , \label{zdist}%
\end{equation}
where we set $b=\hbar=a_{0}=1$. It is plotted in Fig. 2. Since the isotropy of
the ground state, the distributions of the quantity $\mathbf{e}_{r}P_{r}$
along $x$ and $y$-axis are the same as that given by (\ref{zdist}).

Two immediate remarks are necessarily. i) The proper definition of the square
of $\mathbf{e}_{r}P_{r}$ is given by Eq. (\ref{pr2}), and can not be simply
defined as $-\hbar^{2}\nabla_{cart}^{2}-\mathbf{\Pi}^{2}$ because the
noncommutability between $-\hbar^{2}\nabla_{cart}^{2}$ and $\mathbf{\Pi}^{2}$.
These two operators are evidently self-adjoint, making the operator $P_{r}%
^{2}$ (\ref{pr2}) indirectly measurable as well. ii) We note that Dirac had
not attempted to use generalized momentum $P_{\theta}=-i\hbar(\partial
_{\theta}+\cot\theta/2)$, and in our approach, the transverse momentum
(\ref{geom}) suffices to serve any purpose.

\section{Conclusions and discussions}

We in fact accept the Dirac's observation that the \textit{canonical
quantization entails using Cartesian coordinates, }%
\cite{dirac,Schiff,KS,Greiner}\textit{ }and examine the\textit{\ }operator
$\mathbf{e}_{r}P_{r}$ rather than $P_{r}$ itself. Then we demonstrate that
there is a decomposition of $\mathbf{e}_{r}P_{r}$ into a difference of two
self-adjoint operators each having well-defined eigenstates, in which one is
the total momentum and another is the transverse one which is intensively
studied recently as geometric momentum for the two-dimensional spherical
surface, \cite{liu13-1,liu11,liu13-2,133,134,135} etc. \cite{136,137} Explicit
results show the distribution of the $\mathbf{e}_{r}P_{r}$ for the ground
state of the Hydrogen atom can be meaningfully defined. This study supports
Dirac's understanding of the radial momentum operator, which is physically
self-consistent and appealing.

\begin{acknowledgments}
This work is financially supported by National Natural Science Foundation of
China under Grant No. 11175063.
\end{acknowledgments}

\end{document}